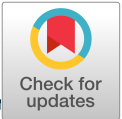

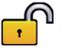

**Correspondence to:**
K. M. Laundal,
karl.laundal@uib.no

# Magnetic Effects of Plasma Pressure Gradients in the Upper $F$ Region

**K. M. Laundal[1]** , **S. M. Hatch[1]** , and **T. Moretto[1]**

[1]Birkeland Centre for Space Science, University of Bergen, Bergen, Norway

**Abstract** The Swarm satellites fly at altitudes that at polar latitudes are generally assumed to only contain currents that are aligned with the local magnetic field. Therefore, disturbances along the main field direction are mainly signatures of auroral electrojet currents, with a relatively smooth structure due to the distance from the currents. Here we show that superimposed on this smooth signal is an irregular pattern of small perturbations, which are anticorrelated with the plasma density measured by the Langmuir probe. We show that the perturbations can be remarkably well reproduced by assuming they represent a $\mathbf{j} \times \mathbf{B}$ force, which balances the plasma pressure gradient implied by the density variations. The associated diamagnetic current, previously reported to be most important near the equator, appears to be a ubiquitous phenomenon also at polar latitudes. A spectral analysis indicates that this effect dominates magnetic field intensity variations at small-scale sizes of a few tens of kilometers.

**Plain Language Summary** The Swarm satellites fly at altitudes that at polar latitudes are generally assumed to only contain currents that are aligned with the local magnetic field. Therefore, disturbances in the magnetic field strength are mainly signatures of horizontal currents below the satellites. Such disturbances have a smooth structure due to the distance from the currents. Here we show that superimposed on this smooth signal is an irregular pattern of small perturbations, which are anticorrelated with the local plasma density. We show that this anticorrelation can be explained in terms of pressure balance between particles and the magnetic field. The electric currents that are associated with the magnetic field fluctuations, diamagnetic currents, have previously been reported to be most important near the equator. Our results show that they are a ubiquitous phenomenon at all latitudes and indicate that the diamagnetic effect dominates magnetic field intensity variations at small-scale sizes of a few tens of kilometers.

## 1. Introduction

In equilibrium, collisionless space plasmas are subject to force balance mainly between particle pressure gradients and magnetic pressure. If the magnetic field is approximately uniform, as is the case in the ionosphere on scale sizes less than ~1,000 km, and the plasma pressure is isotropic, we have

$$\nabla \left( \frac{B^2}{2\mu_0} + p \right) = 0 \Rightarrow \nabla B = -\mu_0 \frac{\nabla p}{B}, \tag{1}$$

where $B$ is the magnetic field strength, $\mu_0$ the vacuum permeability, $p$ the plasma pressure, and we used that $\nabla B^2 = 2B\nabla B$. In this and all subsequent equations, we are only concerned with vector components perpendicular to the magnetic field. Assuming isotropic pressure, local thermodynamic equilibrium, and that variations in pressure are dominated by the density $n$, we use $p = nk_B(T_e + T_i)$ and get

$$\nabla B = -k_B \mu_0 \frac{T_e + T_i}{B} \nabla n, \tag{2}$$

where $k_B$ is Boltzmann's constant and $T_e$ and $T_i$ are the electron and ion temperatures. Equation (2) shows that magnetic field intensity variations in the ionosphere can be produced by variations in plasma density; in regions of increased density, the magnetic field is depressed. The current associated with these magnetic field depressions, so-called diamagnetic currents, have been shown to be a significant source of magnetic field variations in the upper $F$ region at low latitudes (Alken, 2016; Alken et al., 2011, 2017; Lühr et al., 2002, 2003; Stolle et al., 2006). Alken et al. (2011) used global simulation results to test assumptions used by Lühr et al. (2003), who was the first to relate magnetic field perturbations at low Earth orbit to plasma density variations, using the same approach as described here. They found only small discrepancies, possibly due to field line curvature effects, which are neglected in equation (1). Alken et al. (2011) also simulated gravity-induced currents that vary in proportion to the cross product between **B** and the gravitational acceleration **g**. Such currents are important at low latitudes (Alken et al., 2017) but are presumably negligible at high latitudes where **B** and **g** are almost parallel.

To our knowledge, the only previous report of diamagnetic effects at high latitudes was presented by Park et al. (2012). They showed that magnetic field perturbations in the auroral zone, measured with the CHAMP satellite, occasionally were anticorrelated with electron density. In this study they used uncalibrated density measurements from the Digital Ion Drift Meter, taking advantage of the higher sampling rate (1 Hz) compared to the calibrated density measurements from the Planar Langmuir Probe, which had a sampling rate of only 1/15 Hz. With the launch of the Swarm satellite trio in 2013, calibrated simultaneous measurements of the magnetic field, plasma density, and electron temperature (Lomidze et al., 2018) are now available at 2 Hz. In this paper we use these measurements to investigate the relationship between variations in plasma pressure and magnetic field at high latitudes. A case study and statistical analyses are presented in the next section. Section 3 discusses the results, and section 4 concludes the paper.

## 2. Observations

We use the Swarm Alpha (A) and Charlie (C) satellites to approximate components of $\Delta B$ and $\Delta n$. Swarm A and Swarm C fly in nearly the same orbit, separated in the east-west direction by about 1°, with one satellite lagging the other by $\approx 5$ s. In this approximation the vector equation (2) reduces to a scalar equation

$$\Delta B_n = -k_B \mu_0 \frac{2T_e}{B_0} \Delta n,  \tag{3}$$

where $\Delta$ signifies the approximation of the gradient, either along the satellite tracks, with a single satellite approach, or between the two satellites when evaluating interspacecraft differences. Since we only evaluate measurements of the electron temperature, we make the assumption that $T_i = T_e$; hence, the factor of 2. $B_0$ is here specified as the magnetic field strength associated with internal (core and crust) and magnetospheric sources, according to the CHAOS model (Finlay et al., 2015), and $\Delta B$ is estimated from the scalar residuals after subtraction of CHAOS model values.

In the single satellite approach, the original 50-Hz magnetic field residuals are first reduced to the same frequency as the density measurements, 2 Hz, using a 25-point rolling boxcar average. Then, a low-frequency background is subtracted, derived using a second-order Savitzky Golay filter, in 60-s windows. No detrending is applied to the density. The along-track gradient is calculated in 5-s windows by performing a least squares linear fit to the data and taking the slope in units of nanoteslas (or per cubic meter in the case of density) per kilometer. At each time step we use the 5-s mean values of $T_e$ and $B_0$. The window choices essentially imply a band-pass filter, which leaves only variations roughly between 5 and 60 s. This is done for practical rather than physical reasons, since the equations should hold on all scales that are described by magnetohydrodynamics, if steady state is maintained long enough that the observed variations can be interpreted as spatial and not temporal. A spectral analysis of measured and estimated magnetic field fluctuations is presented later in this section.

In the dual-satellite approach $\Delta B$ is calculated by subtracting the detrended CHAOS residuals from Swarm A from Swarm C and dividing by the distance between the two spacecraft. $\Delta n$ is calculated in the same way, but without any prior detrending. The advantage of a dual-satellite approach is that we can relax the assumption about stationarity since the gradient is estimated with simultaneous measurements instead of data collected in a window of 5 s.

Both of these approaches are illustrated in Figure 1, which shows 5 min of observations during a Southern Hemisphere polar cap crossing from dawn to dusk. The upper panel shows the satellite tracks in quasi-dipole latitude and magnetic local time, with the 5-min interval indicated between the triangles. Swarm A is shown in red and Swarm C in blue. The time series shown in red are density and magnetic field measurements from Swarm A. The 5-s moving average electron temperature is shown in gray. The magnetic field (thick red line) is very smooth, as expected in the scalar component, which mostly reflects remote currents (Aakjær et al., 2016; Olsen, 1996). The distance to the current, in this case about 300 km, corresponds approximately to the smallest structure that can be resolved with magnetometers. Detrending and magnifying the magnetic field intensity reveals the presence of the fine structures indicated by the thin red line, which is scaled by a factor of 100 relative to the labels on the $y$ axis and then added to the trend for visualization purposes. Inspection of this curve shows that it closely follows the variations in density, only with opposite polarity. The green and orange lines in the lower panels show the observed $\Delta B$ and estimates $\Delta B_n$ based on equation (3). The three pairs of lines in the lower panel represent the different approaches for estimating the gradients described above. The plasma-based estimates show excellent agreement with the magnetic field measurements in all cases.

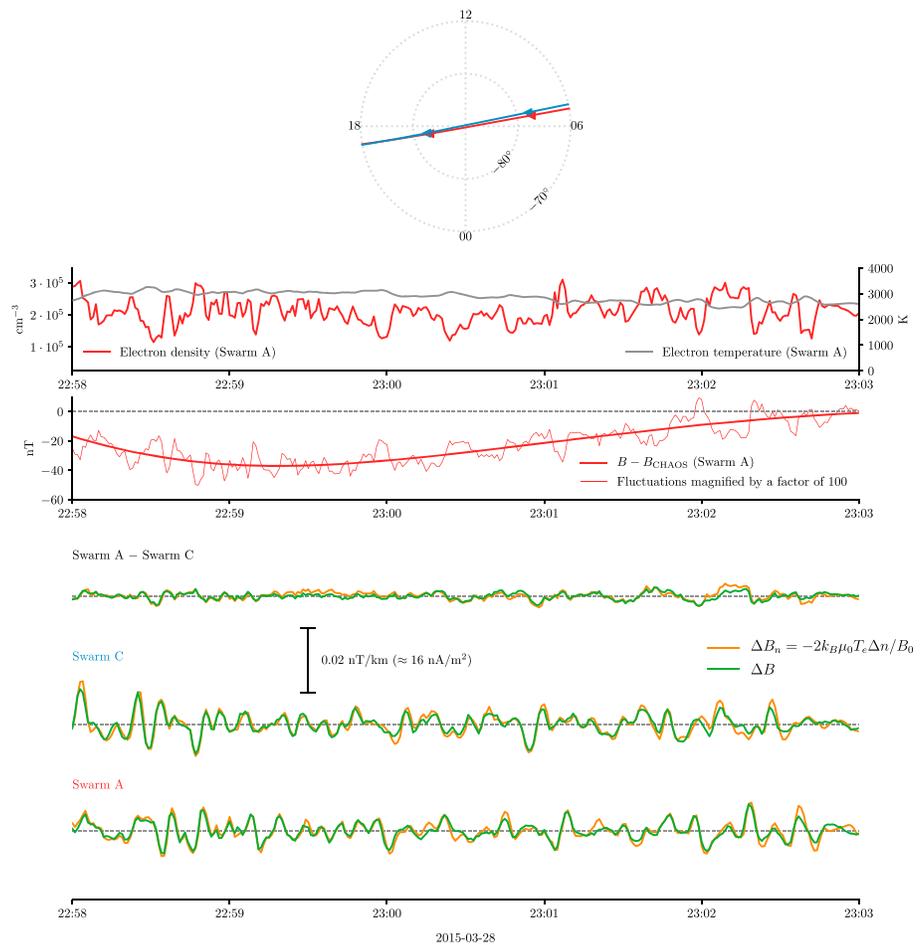

**Figure 1.** A 5-min segment from 28 March 2015, when Swarm A and Swarm C crossed the southern polar cap. The plot limits are indicated by triangles in the satellite tracks in the polar plot, in quasi-dipole latitude and magnetic local time. The two upper time series show plasma density (red) and electron temperature (gray) measured by Swarm A. The next panel shows the CHAOS-corrected magnetic field strength, also from Swarm A. The thin red line shows a representation of the high-frequency magnetic field fluctuations relative to the baseline, exaggerated by a factor of 100. The green and orange curves show the gradient of $B$ (green) and the magnetic field disturbance required to balance the pressure gradient, estimated using plasma and electron temperature measurements (orange). The three pairs correspond to (top) calculating the gradients using interspacecraft differences between Swarm A and Swarm C, and (middle) calculating the gradients using 5 s of data from Swarm C and (bottom) Swarm A.

For each of three pairs of lines the unit of $\Delta B$ and $\Delta B_n$ is given as nanoteslas per kilometer. With the 5-s windows used for gradient estimation, and a satellite speed of 7.5 km/s, the 0.02-nT/km vertical scale bar shown in Figure 1 corresponds to an absolute magnetic field variation of 0.75 nT. Using Ampere's law, the magnetic field gradient can be interpreted as a current density by multiplying by $\mu_0$. The gradients resolved in Figure 1 correspond to current density components, perpendicular to the line separating the measurements, of ~1–10 nA/m². These currents are local, flowing perpendicular to the magnetic field at satellite altitudes. They are at least 1 order of magnitude less than large-scale average field-aligned currents (e.g., Laundal et al., 2018). They are similar to or slightly weaker than diamagnetic currents observed at low latitudes (Alken et al., 2017).

The gradients estimated with the dual spacecraft technique, the top pair of curves in the bottom panel of Figure 1, are generally smaller than the gradients estimated with single spacecraft estimates. The reasons for this probably have to do with differences in measurement geometry. In the beginning of the pass, the line separating the two satellites is largely east-west, while the satellite track itself is north-south. If plasma density structures are largely east-west aligned, which is often the case with polar geospace phenomena, larger gradients are expected in the north-south direction. This idea is supported by the increasing dual spacecraft gradient estimates toward the end of the pass, when the satellite orbits cross. However, even though the gradient components are in the same direction, the dual-satellite estimates remain smaller than the single spacecraft approach. The remaining difference may be explained by the time lag between the satellites at this time, which was 8 s. This means that the line separating the points used to estimate the gradient are about 50% longer than with the 5-s intervals used in the single satellite approach. Sharp gradients in density will therefore appear smaller by a similar factor, consistent with the observations in Figure 1.

The Figure 1 example of almost perfectly matching variations in pressure and magnetic field strength is particularly long lasting, but not unique. Figure 2 shows results of a statistical analysis to investigate when and where the magnetic field variations are well explained by plasma pressure. The analysis is based on all the data from 2015 to 2017 from Swarm A, processed in the same way as the data used in Figure 1. The polar maps are based on the Pearson correlation coefficient between the observed and inferred magnetic field variations, calculated in 20-s windows with no overlap. An occurrence probability is calculated in bins by dividing the number of events in each bin for which the correlation exceeded 0.7 by the total number of observations in that bin. The bins are defined in quasi-dipole latitude and magnetic local time (Richmond, 1995), calculated using the software by Emmert et al. (2010) and the Python wrapper by van der Meeren et al. (2018).

The top row shows the spatial distribution of occurrence probability in the Northern and Southern Hemispheres for the entire 2015–2017 data set. The maps show highest probability in the polar cap, presumably because of the frequent occurrence of polar cap patches (e.g., Spicher et al., 2017), which are localized regions of enhanced plasma density, that are produced by solar extreme ultraviolet radiation on the dayside and transported into the polar cap by intermittent Dungey cycle convection. While polar cap patches are specific plasma structures that by definition are limited to the polar cap, the occurrence statistics capture variations associated with any type of pressure gradient. Thus, the occurrence probability is significant also in the auroral zone and equatorward of the auroral zone on the nightside.

In the second row we combine data from the two hemispheres but divide the data according to seasons, in 90-day bins centered at the solstices. The maps show significantly higher probabilities in summer than winter, the opposite of what was reported by Park et al. (2012). In the bottom set of maps we divide the data according to geomagnetic activity via the *Ap* index. These maps show that the occurrence probability in the polar cap is much higher during active times than during quiet times, while in the auroral zone the probabilities are similar. Equatorward of the auroral zone, the probability is again highest during active times.

The polar maps in Figure 2 suggest that the pressure effect on the magnetic field is almost always present, but only detectable with our technique when pressure gradients are prevalent and other magnetic field disturbances relatively low. These conclusions are based on (1) the peak occurrence in the polar cap, where other magnetic disturbances are low compared to inside the auroral oval, (2) the higher occurrence in summer than winter, when more plasma is produced by sunlight and density gradients are larger, and (3) the higher occurrence for high geomagnetic activity than low activity, probably because of plasma production by precipitation and by structured convection, which creates plasma density gradients. Low occurrence rates do not necessarily mean that the magnetic field is unaffected by plasma pressure; the effect can be uniform on the spatial scales that we investigate, which leads to low covariance. We test this idea in the bottom plot in Figure 2, which shows the median correlation coefficients in bins defined by the amplitude of $\Delta B_n$, which is inferred from density variations. The figure is based on all data from latitudes poleward of $\pm50°$. It shows that when the density variations are large, the correlation increases.

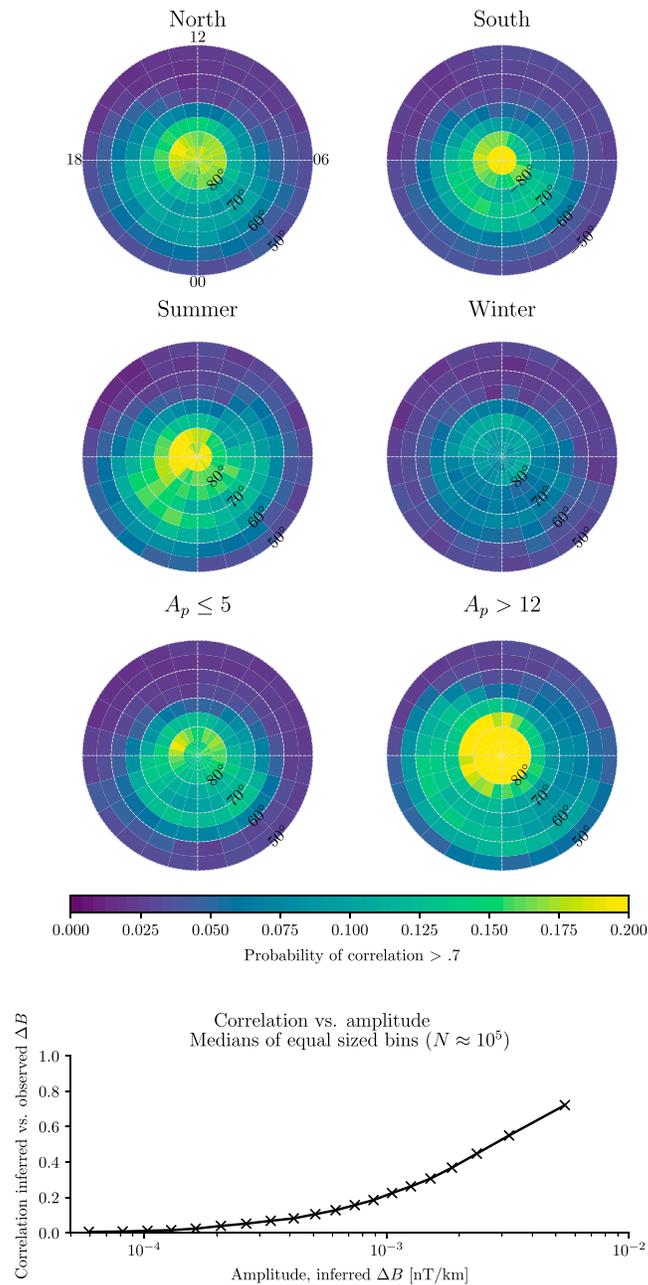

**Figure 2.** Climatological analysis of the correspondence between density and magnetic field variations, based on all available data from Swarm A in 2015. Each polar map shows the occurrence probability of the correlation between observed and density-inferred magnetic field variations in 20-s windows being greater than 0.7. The top maps show the occurrence probabilities in the Northern and Southern Hemispheres. In the second and third rows, the hemispheres are combined, but data selected according to seasons, 90-day intervals centered at the solstices, or geomagnetic activity level, defined by the $Ap$ index. The bottom plot shows the median correlation coefficient versus the median amplitude, in equal-sized bins defined by the amplitude. The amplitude is defined as the maximum minus minimum in the same 20-s windows used to calculate the correlation coefficients.

Xiong et al. (2018), in a study that focused on the relationship between density gradients and the loss of GPS signals, presented climatological maps of density variations along the Swarm C satellite track. The average density gradients during orbits when GPS signal was lost resembles the occurrence patterns in Figure 2, with one notable exception: Xiong et al. (2018) reported a clear peak toward the dayside, near the cusp, while the peak occurrence in the maps of Figure 2 is well inside the polar cap. The cusp is a region where plasma is being produced by precipitation and sunlight, and it may not have reached equilibrium. Inertial effects, large variations in plasma temperature, and large differences between electron and ion temperatures violate the assumptions used to derive equation (3). This may reduce the occurrence rates in Figure 2 despite the presence of large density gradients. Later, when the plasma structures have moved into the polar cap, equilibrium has been reached and the magnetic field and plasma density become well correlated.

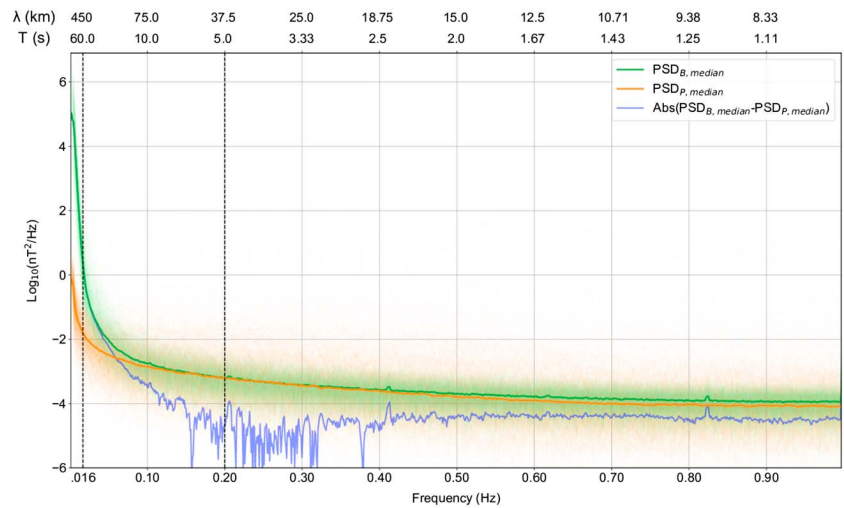

**Figure 3.** Median power spectral density (PSD) of the scalar magnetic field $PSD_B$ (green line), corrected for main field contributions using the CHAOS model, the estimated magnetic field disturbance due to pressure balance $PSD_P$ (orange line), and the absolute difference $PSD_B - PSD_P$ (blue line) for $N = 1,706$ polar cap/auroral zone crossings made by Swarm A. The PSDs from each individual crossing are shown with thin lines in the background. The dashed lines indicate approximately the frequency band that is analyzed in Figures 1 and 2. The period and wavelength indicated in the top $x$ axis are calculated as $1/f_{sc}$ and $v_s/f_{sc}$, respectively, where $f_{sc}$ is the frequency in the lower $x$ axis, and $v_s = 7.5$ km/s is approximately the satellite speed.

The results presented above are valid for a specific range of spatial scales. Assuming the observed variations are purely due to the spacecraft traversing static density structures (i.e., spatial Doppler-shifted structures) and a satellite speed of $\approx 7.5$ km/s, the 5- and 60-s windows used for differentiation and detrending, respectively, correspond to a range of spatial scales between approximately 40 and 400 km. Variations over a larger range of spatial scales are displayed in Figure 3, which presents the power spectral densities (PSDs) of $B - B_{CHAOS}$ (green lines) and of pressure-induced magnetic field disturbances (orange lines) as functions of spacecraft-frame frequency $f_{sc}$. Spacecraft-frame frequencies refer to the variations measured by the spacecraft, either because of time variations in the magnetic field or density or because the spacecraft traverses spatial structures. We adopt the latter interpretation below, although the two effects cannot be distinguished with a single spacecraft. While the frequency is shown in the lower $x$ axis, the upper $x$ axis shows the corresponding wave period, $1/f_{sc}$, and spatial scale, $v_s/f_{sc}$, where the spacecraft speed $v_s$ is set to 7.5 km/s.

Instead of the gradient, the orange lines in Figure 3 are based on the absolute pressure-induced disturbance, calculated as $k_B \mu_0 2 T_e n_e / B_0$ (Alken et al., 2017; Lühr et al., 2003). The thick lines represent the median power spectrum density from 1,706 individual PSDs (shown as thin lines in the background). Each PSD is estimated via the multitaper method (Hatch & LaBelle, 2017; Slepian, 1978; Thomson, 1982) from a 16-min time series that is centered on the highest magnetic latitude reached by Swarm A during that crossing. The absolute difference between the median PSDs is indicated by the blue line.

Figure 3 shows that the wave power increases with decreasing frequency or increasing spatial scale, for both the measured magnetic field and the plasma-derived magnetic field disturbance. The plasma-derived magnetic field approaches 1 nT²/Hz (0 on the log scale), which corresponds to about 1-nT amplitudes. At these low frequencies, the measured magnetic field exceeds the plasma-derived magnetic field by several orders of magnitude. The main contribution to the magnetic field at these frequencies is probably the auroral electrojets. Since the electrojet is about 300 km below the satellite, the signal drops off sharply at similar horizontal spatial scales and becomes comparable in magnitude to the plasma-derived magnetic perturbations. The dashed vertical lines indicate roughly the frequency range that is investigated in Figures 1 and 2. Within this frequency band, electrojet signals can be significant, which is probably why the occurrence probability in Figure 2 is relatively low in the auroral zone, where the electrojet is strong. The occurrence probability is higher away from the auroral zone, where only large-scale signatures of the electrojet are present, but removed by high-pass filtering.

The magnitudes of the measured and plasma-derived magnetic field wave power in Figure 3 remain similar as the frequency increases to the highest frequency that we can resolve, 1 Hz. This leads us to conclude that at small scales, smaller than a few tens of kilometers, plasma pressure variations are the dominating source of variations in the magnetic field strength. Other types of small-scale fluctuations, like Pc1–Pc2 waves, which are not directly related to local pressure variations, do not appear to have a strong effect on the median magnetic field PSD. Field-aligned currents, which exist at all scales investigated here (e.g., Gjerloev et al., 2011), are only associated with changes in the magnetic field orientation and not its magnitude.

## 3. Discussion

We have shown that plasma pressure variations at polar latitudes are associated with small but clearly identifiable variations in the magnetic field strength at Swarm altitude, approximately 450 km. The variations in magnetic field strength can in many cases be precisely estimated using the concept of force balance together with measurements of the plasma density and electron temperature. This phenomenon has been extensively studied at low latitudes (Alken et al., 2011, 2017; Lühr et al., 2003), where so-called diamagnetic disturbances can reach 5 nT. At polar latitudes, only Park et al. (2012) have reported observations of the diamagnetic effect, based on data from the CHAMP satellite. Our study is the first to investigate the diamagnetic effect at polar latitudes using the Swarm satellites, which provide more precise measurements, at higher cadence, than earlier missions.

There are many important differences between our findings and the results by Park et al. (2012). First, their occurrence probabilities were about 1 order of magnitude less than what we find. This may be explained by the higher sensitivity of the Swarm instruments but could also be related to methodological differences: Park et al. (2012) based their analysis on high-pass-filtered magnetic field and density measurements directly, and not estimates of their spatial gradients, as we have done here. It is therefore more significant that the distribution of events is also very different: Park et al. (2012) reported highest occurrence rates in the auroral oval, while we have a clear peak in the polar cap. It is possible that this is because their study only included very strong perturbations, while our Swarm-based analysis resolves finer variations.

Park et al. (2012) also reported higher occurrence rates in winter than summer, opposite of what we find. The contradicting results may be related to the rather complex seasonal variation of polar cap patches, discussed by, for example, Chartier et al. (2018). While some studies of polar cap patches find that they are most frequent during local winter (e.g., Spicher et al., 2017), others report that the occurrence rate peaks in December for both hemispheres. According to Chartier et al. (2018), the key reason for the different conclusions is the polar cap patch detection algorithm: If only relative changes in plasma density are considered, small density increases in winter will be counted while quite large increases in summer are missed. Furthermore, the background density will be different in the two hemispheres due to differences in solar illumination relative to the magnetic field (e.g., Laundal et al., 2017). Even though polar cap patches are probably a significant contribution to the variations reported in this paper, it is not clear exactly how their seasonal variation influences our results. Nevertheless, these unresolved issues suggest that the seasonal and hemispheric statistics that are reported in Figure 2 may conceal more complex variations in density-induced magnetic field variations that should be addressed in future studies. While 3 years of data was enough to give well-defined statistics in Figure 2, a detailed study of the annual variations will require a larger data set. This can be provided by the Swarm satellites, launched in late 2013 and still in operation.

### 3.1. Potential Applications

There are several potential applications of the results of the present paper, which could be explored in future work. First, it shows that density and temperature measurements could be used to correct magnetic field measurements for the diamagnetic effect, similar to what was suggested by Lühr et al. (2003). This could be useful in modeling of the main magnetic field, which usually only considers the scalar field at high latitudes to avoid disturbances related to field-aligned currents (Finlay et al., 2017). Despite this precaution, and strict selection criteria to avoid disturbance fields, clear signatures of ionospheric currents are seen in the data (Friis-Christensen et al., 2017). The diamagnetic disturbances discussed here, though weak, may pose an additional challenge since they seem to appear preferentially at the same geographic locations (polar caps) and vary on scales that modern main field models aim to resolve. Figure 3 shows that on large scales, or low frequencies, the median pressure-induced perturbation approaches 0 on the log scale, which corresponds to an amplitude of ~1 nT. As shown by Lühr et al. (2003), and in this paper, these disturbances can potentially be corrected using density and temperature measurements.

Another, perhaps more speculative, application is to estimate the ion temperature. One of the approximations that we have made is that the electron and ion temperatures are equal. Equation (2) can be used to express the ion temperature in terms of the magnetic field gradient, plasma density gradient, and electron temperature:

$$T_i = -\frac{\Delta B}{\Delta n}\frac{B_0}{k_B \mu_0} - T_e = T_e\left(2\frac{\Delta B}{\Delta B_n} - 1\right), \tag{4}$$

where $\Delta B_n$ is the magnetic perturbation estimated with equation (3), which uses plasma density and electron temperature measurements, and the assumption that $T_i = T_e$. This approach can be tested by comparison with ion temperature estimates from the thermal ion imager data on the Swarm spacecraft (Knudsen et al., 2017).

Another possible application is to derive along-track plasma convection speed $V$ from cross-correlating magnetic field structures on Swarm A and Swarm C. Assuming that plasma structures remain static during the time it takes for the two satellites to cross them, the plasma motion can be derived by comparing the observed delay $\Delta\tau$, found from cross correlation, to the actual time lag between the two satellites $\Delta t$:

$$V = \frac{\Delta\tau - \Delta t}{\Delta\tau}v_s, \tag{5}$$

where $v_s$ is the speed of the satellites and $V$ is positive when the convection is in the same direction as the satellites. Park et al. (2015) demonstrated this principle using Swarm density measurements from the first few months of the mission, when all three satellites were in a pearl-on-a-string configuration with $\Delta t$ between 1 and 3 min. They noted that with $\Delta t \approx 5$ s, which is the case in the current constellation, the 2-Hz sampling rate of plasma density is too low to give precise estimates of the convection. This is because the lowest possible value in the numerator of equation (5) is 0.5 s, which corresponds to a resolution for $V$ of ~500 m/s when realistic satellite speed and separation time are considered. If the same approach can be applied with the 50-Hz magnetic field measurements, assuming that variations are signatures of static plasma structures, the resolution can be improved to ~20 m/s. However, the assumption of static density structures at these scales is questionable and remains to be demonstrated.

Finally, there may also be applications that are relevant to space weather issues. If fluctuations in magnetic field intensity are due to plasma density variations, the 50-Hz data from Swarm can be used to resolve smaller plasma structures than what is possible with the 2-Hz electron density measurements. This idea was applied with CHAMP magnetic field measurements, by Stolle et al. (2006). Since ionospheric irregularities are associated with the loss of GPS signal (e.g., Xiong et al., 2018), the high-frequency magnetic field measurements may help in increasing our understanding of this natural hazard phenomenon.

## 4. Conclusions

We have shown that plasma pressure variations at polar latitudes in the upper $F$ region, at ≈450-km altitude, are associated with variations in the magnetic field intensity. These magnetic field variations can be precisely estimated by assuming that the magnetic pressure and particle pressure balance. High correlations between pressure-derived and actual magnetic field gradients are found most frequently in the polar cap. High correlations are also found more often in local summer than in winter and more often when the geomagnetic activity is high than when it is low. Our results indicate that the pressure-induced magnetic field variations are present at all spatial scales, and Figure 3 suggests that they may be the dominating cause for such variations at spatial scales of a few tens of kilometers.

**Acknowledgments**
ESA is thanked for providing prompt access to the Swarm L1b data, accessible at http://earth.esa.int/swarm. Swarm CDF files were read in Python using software by Stansby et al. (2018). The *Ap* index is provided by the Helmholtz Centre Potsdam, GFZ German Research Centre For Geosciences, and was downloaded from the website (ftp://ftp.ngdc.noaa.gov/STP/GEOMA GNETIC_DATA/INDICES/KP_AP). CHAOS magnetic field values were calculated with the ChaosMagPy software, available at the website (https://pypi.org/project/chaosmagpy/). The study was funded by the Research Council of Norway/CoE under contract 223252/F50. We thank the two reviewers for very useful comments and suggestions.

## References

Aakjær, C. D., Olsen, N., & Finlay, C. C. (2016). Determining polar ionospheric electrojet currents from Swarm satellite constellation magnetic data. *Earth Planets Space*, 68, 140. https://doi.org/10.1186/s40623-016-0509-y

Alken, P. (2016). Observations and modeling of the ionospheric gravity and diamagnetic current systems from CHAMP and Swarm measurements. *Journal of Geophysical Research: Space Physics*, 121, 589–601. https://doi.org/10.1002/2015JA022163

Alken, P., Maus, S., Richmond, A. D., & Maute, A. (2011). The ionospheric gravity and diamagnetic current systems. *Journal of Geophysical Research*, 116, A12316. https://doi.org/10.1029/2011JA017126

Alken, P., Maute, A., & Richmond, A. D. (2017). The *F*-region gravity and pressure gradient current systems: A review. *Space Science Review*, 206, 451–469. https://doi.org/10.1007/s11214-016-0266-z

Chartier, A. T., Mitchell, C. N., & Miller, E. S. (2018). Annual occurrence rates of ionospheric polar cap patches observed using Swarm. *Journal of Geophysical Research: Space Physics*, 123, 2327–2335. https://doi.org/10.1002/2017JA024811

Emmert, J. T., Richmond, A. D., & Drob, D. P. (2010). A computationally compact representation of Magnetic Apex and Quasi Dipole coordinates with smooth base vectors. *Journal of Geophysical Research*, 115, A08322. https://doi.org/10.1029/2010JA015326

Finlay, C. C., Lesur, V., Thébault, E., Vervelidou, F., Morschhauser, A., & Shore, R. (2017). Challenges handling magnetospheric and ionospheric signals in internal geomagnetic field modelling. *Space Science Reviews*, 206. https://doi.org/10.1007/s11214-016-0285-9

Finlay, C. C., Olsen, N., & Tøffner-Clausen, L. (2015). DTU candidate field models for IGRF-12 and the CHAOS-5 geomagnetic field model. *Earth, Planets and Space*, 67, 157–189. https://doi.org/10.1186/ s40623-015-0274-3

Friis-Christensen, E., Finlay, C. C., Hesse, M., & Laundal, K. M. (2017). Magnetic field perturbations from currents in the dark polar regions during quiet geomagnetic conditions. *Space Science Review*, 206, 281–297. https://doi.org/10.1007/s11214-017-0332-1

Gjerloev, J. W., Ohtani, S., Iijima, T., Anderson, B., Slavin, J., & Le, G. (2011). Characteristics of the terrestrial field-aligned current system. *Annales Geophysicae*, 9(10), 1713–1729. https://doi.org/10.5194/angeo-29-1713-2011

Hatch, S. M., & LaBelle, J. W. (2017). Application of a new method for calculation of low-frequency wave vectors. In *Proceedings of the 8th International Workshop on Planetary, Solar and Heliospheric Radio Emissions held at Seggauberg near Graz, Austria, October 25–27, 2016.*

Knudsen, D. J., Burchill, J. K., Buchert, S. C., Eriksson, A. I., Gill, R., Wahlund, J. E., & Moffat, B. (2017). Thermal ion imagers and Langmuir probes in the Swarm electric field instruments. *Journal of Geophysical Research: Space Physics*, 122, 2655–2673. https://doi.org/10.1002/2016JA022571

Laundal, K. M., Cnossen, I., Milan, S. E., Haaland, S. E., Coxon, J., Pedatella, N. M., & Reistad, J. P. (2017). North-South asymmetries in Earth's magnetic field—Effects on high-latitude geospace. *Space Science Review*, 206, 225–257. https://doi.org/10.1007/s11214-016-0273-0

Laundal, K. M., Finlay, C. C., Olsen, N., & Reistad, J. P. (2018). Solar wind and seasonal influence on ionospheric currents from Swarm and CHAMP measurements. *Journal Geophysical Research: Space Physics*, 123, 4402–4429. https://doi.org/10.1029/2018JA025387

Lomidze, L., Knudsen, D. J., Burchill, J., Kouznetsov, A., & Buchert, S. C. (2018). Calibration and validation of Swarm plasma densities and electron temperatures using ground-based radars and satellite radio occultation measurements. *Radio Science*, 53, 15–36. https://doi.org/10.1002/2017RS006415

Lühr, H., Maus, S., Rother, M., & Cooke, D. (2002). First in situ observation of night-time *F* region currents with the CHAMP satellite. *Geophysical Research Letters*, 29, (10), 1489. https://doi.org/10.1029/2001GL013845

Lühr, H., Rother, M., Maus, S., Mai, W., & Cooke, D. (2003). The diamagnetic effect of the equatorial Appleton anomaly: Its characteristics and impact on geomagnetic field modeling. *Geophysical Research Letters*, 30(17), 1906. https://doi.org/10.1029/2003GL017407

Olsen, N. (1996). A new tool for determining ionospheric currents from magnetic satellite data. *Geophysical Research Letters*, 23(24), 3635–3638. https://doi.org/10.1029/96GL02896

Park, J., Ehrlich, R., Lühr, H., & Ritter, P. (2012). Plasma irregularities in the high-latitude ionospheric *F*-region and their diamagnetic signatures as observed by CHAMP. *Journal of Geophysical Research*, 117, A10. https://doi.org/10.1029/2012JA018166

Park, J., Lühr, H., Stolle, C., Malhotra, G., Baker, J. B. H., Buchert, S., & Gill, R. (2015). Estimating along-track plasma drift speed from electron density measurements by the three Swarm satellites. *Annales Geophysicae*, 33(7), 829–835. https://doi.org/10.5194/angeo-33-829-2015

Richmond, A. D. (1995). Ionospheric electrodynamics using magnetic apex coordinates. *Journal of Geomagnetism and Geoelectricity*, 47, 191–212.

Slepian, D. (1978). Prolate spheroidal wave functions, Fourier analysis, and uncertainty—V: The discrete case. *The Bell System Technical Journal*, 57(5), 1371–1430. https://doi.org/10.1002/j.1538-7305.1978.tb02104.x

Spicher, A., Clausen, L. B. N., Miloch, W. J., Lofstad, V., Jin, Y., & Moen, J. I. (2017). Interhemispheric study of polar cap patch occurrence based on Swarm in situ data. *Journal of Geophysical Research: Space Physics*, 122, 3837–3851. https://doi.org/10.1002/2016JA023750

Stansby, D., Hirsch, M., Harter, B., & Ireland, J. (2018). MAVENSDC/cdflib: CDFLib 0.3.9 (Version 0.3.9). Zenodo. https://doi.org/10.5281/zenodo.2543499

Stolle, C., Lühr, H., Rother, M., & Balasis, G. (2006). Magnetic signatures of equatorial spread *F* as observed by the CHAMP satellite. *Journal of Geophysical Research*, 111, A2. https://doi.org/10.1029/2005JA011184

Thomson, D. J. (1982). Spectrum estimation and harmonic analysis. *Proceedings of the IEEE*, 70(9), 1055–1096. https://doi.org/10.1109/PROC.1982.12433

van der Meeren, C., Burrell, A. G., & Laundal, K. M. (2018). apexpy: ApexPy version 1.0.3. http://doi.org/10.5281/zenodo.1214407

Xiong, C., Stolle, C., & Park, J. (2018). Climatology of GPS signal loss observed by Swarm satellites. *Annales Geophysicae*, 36(2), 679–693. http://doi.org/10.5194/angeo-36-679-2018